# Optimizing Higher-Order Lagrangian Perturbation Theory for Standard CDM and BSI models

Arno G. Weiß[1], Stefan Gottlöber[2], Thomas Buchert[3]

[1] Max-Planck-Institut für Astrophysik, Karl-Schwarzschild-Straße 1, D-85748 Garching, Germany (aow@mpa-garching.mpg.de)

[2] Astrophysikalisches Institut Potsdam, An der Sternwarte 16, D-14482 Potsdam, Germany (sgottloeber@aip.de)

[3] Institut für Theoretische Physik, Ludwig-Maximilians-Universität, Theresienstraße 37, D-80333 München, Germany (buchert@stat.physik.uni-muenchen.de)

# Optimizing Higher-Order Lagrangian Perturbation Theory

# for Standard CDM and BSI models

Arno G. Weiß    Stefan Gottlöber    Thomas Buchert

May 17, 1995

**Abstract**

We investigate the performance of Lagrangian perturbation theory up to the second order for two scenarios of cosmological large-scale structure formation, SCDM (standard cold dark matter) and BSI (broken scale invariance). The latter model we study as a representative of COBE-normalized CDM models which fit the small-scale power of galaxy surveys. In this context we optimize the performance of the Lagrangian perturbation schemes by smoothing the small-scale fluctuations in the initial data. The results of the so obtained Lagrangian mappings are computed for a set of COBE-normalized SCDM and BSI initial data of different sizes and at different times. We compare these results against those obtained with a numerical PM-code. We find an excellent performance of the optimized Lagrangian schemes down to scales close to the correlation length. This is explained by the counterintuitive fact that nonlinearities in the model can produce more small-scale power, if initially such power is removed. The optimization scheme can be expressed in a way which is independent of the type of fluctuation spectrum and of the size of the simulations.

# 1  Introduction

Zel'dovich's celebrated approximation (Zel'dovich 1970) has largely improved the qualitative understanding of how large–scale structure in the Universe forms. Why this approximation is so useful can be traced back to the fact that it can be understood as a subset of the first–order solution in a Lagrangian perturbation approach (Buchert 1989, 1992), which allows following the evolution of gravitational instabilities into the weakly non–linear regime of structure formation, i.e., where the r.m.s. fluctuation of the density contrast is of order unity. This value roughly corresponds to the limit of application of Lagrangian perturbation theory at *any* order. The quantitative demonstration of this limitation in comparison with N–body simulations actually restricts its application to so–called "pancake models" (Buchert *et al.* 1994) which lack power on small scales similar to the traditional Hot–Dark–Matter picture of structure formation. The "Zel'dovich–approximation" has been applied only to initialize N–body codes, or to model structures at mildly non–linear stages down to scales of $10^{15} M_\odot$.

Meanwhile, Zel'dovich's original model has developed into a powerful tool for the modeling of structure formation also on smaller spatial scales with the implication that models with substantial amount of small–scale power can be accessed. The basic idea of improvement relies on the smoothing of small–scale power in the initial conditions, i.e., the main failure of the model to describe the gravitational collapse after shell–crossing is effectively compensated. Recent work (see: Coles *et al.* 1993, Melott *et al.* 1994, and the summary by Melott (1994 and ref. therein), Bouchet *et al.* 1995 and Sathyaprakash *et al.* 1995) has exploited this idea with the result that the first–order Lagrangian perturbation scheme in a truncated form is capable of modeling structure formation down to galaxy group mass–scales ($10^{13} M_\odot$) in any model including hierarchical cosmogonies at fully developed non–linear stages. Going to higher–order Lagrangian perturbation schemes improves the performance of such models only up to the second order approximation, which appears to be remarkably robust and consistently better than the first–order scheme as measured by various statistics (Melott *et al.* 1995).

The present paper is motivated by the need to apply this tool to realistic dark matter models, since earlier systematic analyses were limited to case studies, e.g., power–spectra of power-law form. It is designed to uncover also universal properties of truncation of initial data, i.e. rules which may apply to any initial data set. For example, a Gaussian type of smoothing window and a truncation slightly below the nonlinearity scale was suggested by the earlier systematic studies and will be confirmed for the dark matter spectra analyzed in this work. Therefore, this paper forms the final step to achieve a reliable modeling of large–scale structure with simple analytical tools, which can guide and meet the contemporary efforts of understanding the formation of galaxies in the large–scale environment. Here, we emphasize that a statistically 'fair' domain of the Universe can be modeled fast, effectively and reliable down to those scales which – for a large realization – coincide with the Nyquist frequency of standard N–body runs. There is no need for N–body computing on these scales, while smaller scales are no longer ruled by the gravitational interaction alone, and must be accessed by more complex methods such as hydrodynamical simulations.

The second–order scheme employed in this work is as easy to implement as the first–order scheme. The CPU times on a CRAY YMP are for the first–order scheme 25 seconds, and for the second–order scheme 60 seconds; the corresponding CPU times on a CONVEX C220 are 2 and 5 minutes, respectively. Thus, even the second–order scheme is competitive with one step in a corresponding PM–type N–body simulation.

The paper is organized as follows. In Section 2 we describe the Lagrangian perturbation theory up to the second order and present the optimizing technique used. In Section 3 we discuss N-body simulations of large scale structure formation in a universe dominated by cold dark matter. We use both the primordial Harrison-Zeldovich perturbation spectrum and a primordial spectrum with broken scale invariance (BSI). The latter was introduced (Gottlöber et al. 1991, 1994) in order to solve a discrepancy in the COBE normalized SCDM model which shows too much structure on small scales. Both in the SCDM and BSI model the universe is dominated by cold dark matter. In Section 4 we present both the results of our optimizing procedure and compare the performance of the optimized Lagrangian perturbation theory against our numerical simulations. In Section 5 we present our conclusions from these results. Throughout this paper we assume that the Hubble constant is 50 km/s/Mpc, however, for comparison with other works, we present our results in terms of an Hubble constant $H = h100$ km/s/Mpc with $h = 1/2$.

## 2  Lagrangian Theory and Optimization

The Lagrangian perturbation schemes which we employ in the present work together with a description of their realization can be found in (Buchert 1993, 1994, Buchert et al. 1994, Melott et al. 1995). Like in the latter work we restrict the presentation of results here to the first– and second–order schemes, the second–order scheme is expected to contain the major effects on large scales according to the results of Buchert et al. (1994) for the present purpose.

Denoting comoving Eulerian coordinates by $\mathbf{q}$ and Lagrangian coordinates by $\mathbf{X}$, the field of trajectories $\mathbf{q} = \mathbf{F}(\mathbf{X}, t)$ up to the second order on an Enstein–de Sitter background reads:

$$\mathbf{F} = \mathbf{X} + q_1(a)\, \nabla_0 \mathcal{S}^{(1)}(\mathbf{X}) + q_2(a)\, \nabla_0 \mathcal{S}^{(2)}(\mathbf{X}), \tag{1}$$

with the time–dependent coefficients expressed in terms of the expansion function $a(t) = (\frac{t}{t_i})^{2/3}$:

$$q_1 = \left(\frac{3}{2}\right)(a-1), \tag{2}$$

$$q_2 = \left(\frac{3}{2}\right)^2 \left(-\frac{3}{14}a^2 + \frac{3}{5}a - \frac{1}{2} + \frac{4}{35}a^{-\frac{3}{2}}\right). \tag{3}$$

The perturbation potentials have to be constructed by solving iteratively the two nonlocal boundary value problems:

$$\Delta_0 \mathcal{S}^{(1)} = I(\mathcal{S}_{,i,k})t_i, \tag{4}$$

$$\Delta_0 \mathcal{S}^{(2)} = 2II(\mathcal{S}^{(1)}_{,i,k}), \tag{5}$$

where $I$ and $II$ denote the first and second principal scalar invariants of the tensor gradient $(\mathcal{S}^{(1)}_{,i,k})$:

$$I(\mathcal{S}^{(1)}_{,i,k}) = \text{tr}(\mathcal{S}^{(1)}_{,i,k}) = \Delta_0 \mathcal{S}^{(1)}, \tag{6}$$

$$II(\mathcal{S}^{(1)}_{,i,k}) = \frac{1}{2}[(\text{tr}(\mathcal{S}^{(1)}_{,i,k}))^2 - \text{tr}((\mathcal{S}^{(1)}_{,i,k})^2)]. \tag{7}$$

The general set of initial data is restricted according to the assumption of parallelism of the peculiar–velocity field and the peculiar–acceleration field at the initial time $t_i$ (see Buchert 1994 and ref. therein for a discussion of this restriction). Therefore, the initial fluctuation field can be specified to the peculiar–velocity potential $\mathcal{S}$ alone. We can set $\mathcal{S}^{(1)} = \mathcal{S}t_i$ (which is the unique solution of the first Poisson equation (4) for periodic initial data, see Buchert 1994, Appendix B). Doing this, the flow–field (1) reduces to Zel'dovich's approximation if restricted to the first order.

We realize the solution by first solving Poisson's equation for $\mathcal{S}$ via FFT (Fast Fourier Transform) from the initial density contrast $\delta$ generated as initial data for the numerical simulation. In an Einstein–de Sitter model we have:

$$\Delta_0 \mathcal{S} = -\frac{2}{3t_i}\delta. \tag{8}$$

We then calculate the second principal invariant $II$ directly from $\mathcal{S}$ and solve the second Poisson equation (7) using FFT. The density in the analytical models is calculated by collecting trajectories of the Lagrangian perturbation solutions at the different orders into a $64^3$ pixel grid with the same method (CIC binning) as in the N–body simulation.

The optimization of these approximations is done as proposed by Coles et al. (1993) by smoothing the high-frequency part of the Fourier transform of the initial density field.

In previous work (Melott *et al.* 1995) we found that convolving the initial density field with a Gaussian consistently gives the best results for both first- and second-order. The characteristic smoothing scale can best be given as the width $k_{gs}$ of the Gaussian window-function in $k$-space

$$\tilde{W}_{gs}(k) = e^{-\frac{k^2}{2k_{gs}^2}} \qquad (9)$$

in units of the scale of nonlinearity $k_{nl}$, defined as the scale on which the integral of the linearly evolved power-spectrum of the initial data becomes unity:

$$\frac{a^2(t)}{(2\pi)^3} \int_0^{k_{\rm nl}} d^3k\, P(k) = 1. \qquad (10)$$

The error which is introduced here by setting the lower bound in the calculation of this integral to the fundamental wave number $k_f$ of our simulation is negligible for box-sizes of the simulations considered here (i.e., the error in estimating $k_{nl}$ is less than one grid unit of our spatial resolution).

Here we investigate the dependence of the optimal smoothing on the initial dark matter power-spectrum, on the size of the simulation and on the final time by conducting a "loop" over the smoothing scale $k_{gs}$ for the different models, box-sizes and times. We run both the numerical simulation and the analytical approximation schemes with a resolution of $128^3$ particles, which we collect at the final stages into a density grid of $64^3$ cells, using the clouds-in-cells algorithm. A comparison of the density fields obtained by the different algorithms is then done by evaluating the cross-correlation coefficient

$$S = \frac{\langle \delta_1 \delta_2 \rangle}{\sigma_1 \sigma_2}, \qquad (11)$$

where $\langle \ldots \rangle$ denotes spatial averaging over the whole volume of the simulation, and $\sigma_1$, $\sigma_2$ are the rms-density contrasts of the density contrast fields $\delta_1$ of the numerical simulation and $\delta_2$ of the analytical schemes, respectively. The cross-correlation coefficient $S$ measures whether mass is moved to the right place. Computing $S$ on a cell-to-cell basis for a set of initial data smoothed over a series of smoothing lengths yields the optimum smoothing scale as the one that maximizes $S$. Computing $S$ for final density fields smoothed with a Gaussian we get a scale-dependent quality estimate for our analytical schemes compared against the numerical simulations.

Other statistics we consider here for a comparison of our analytical schemes against the numerical simulations include the power-spectral analysis of the final density fields, the density distribution functions and the two-point correlation functions. For a more detailed discussion of the statistics suitable for a comparison of the density fields obtained by different algorithms see, e.g., Melott *et al.* (1994,1995).

# 3 Numerical simulations with COBE-normalized SCDM and BSI initial data

In order to follow the nonlinear evolution of the formation of structure we have performed N-body calculations using a standard PM code (Kates *et al.* 1991) with $128^3$ particles on a $256^3$ grid (Kates *et al.* 1995). The universe is assumed to be spatially flat ($\Omega = 1$). It is dominated by cold dark matter. We consider here simulations which were performed in boxes of 500 $h^{-1}$ Mpc, 200 $h^{-1}$ Mpc and 75 $h^{-1}$ Mpc. The simulations were started with the power spectrum $P(k)$ of density perturbations calculated at $z = 25$,

$$P(k) = \left(\frac{\delta\rho}{\rho}\right)^2_{\mathbf{k}} = \frac{4}{9}(\frac{kR_h}{2})^4 \Phi^2(k) T^2(k), \qquad (12)$$

where $R_h = 2H^{-1}$ denotes the horizon. The primordial perturbation spectrum $\Phi$ is either the Harrison-Zel'dovich spectrum ($\Phi = const$) of the SCDM model or the spectrum with broken scale invariance calculated from a double inflationary model (Gottlöber *et al.* 1991). In the BSI model the spectrum is of Harrison-Zel'dovich type both in the limit of very large and very small scales.

We have used the CDM transfer function $T(k)$ of Bond and Efstathiou (1984). For all simulations we have normalized our spectra using the 10°-variance of the CMB fluctuations $\sigma_T = (30 \pm 7.5)\mu$K of the first year COBE data (Smoot *et al.* 1992). Thus the power spectrum of the BSI model shows less power on small scales than the SCDM model. The scales of nonlinearity defined in eq.(10) are $\lambda_{nl}^{SCDM} = 27h^{-1}$ Mpc and $\lambda_{nl}^{BSI} = 7h^{-1}$ Mpc. In Fig. 1 we show the linear BSI and SCDM spectra and indicate the box sizes and the resolution of our simulations.

The first and second year COBE data were analyzed by many authors using different statistical techniques. The new normalization of Górski *et al.* (1994) is about 25 % higher than the normalization which we used in the simulations. Consequently, in this normalization the scales of nonlinearity would increase to $\lambda_{nl}^{SCDM} = 33h^{-1}$ Mpc and $\lambda_{nl}^{BSI} = 11h^{-1}$ Mpc.

# 4 Results

In Fig. 2 we have collected the results of our optimization procedure in a scatter plot of $k_{gs}/k_{nl}$ against the effective power index $n(k_{nl})$ on the scale of nonlinearity, which is obtained as the tangent to $P(k)$ in a log-log plot. Interestingly, the optimum smoothing scales for both first- and second-order show no significant variation with $n(k_{nl})$ over the whole range of $n(k_{nl})$ investigated here, ranging from $n = -2.4$ to $n = -0.4$. A linear regression fit of the data shows only a slightly negative slope of $k_{gs}/k_{nl}$ with respect to $n(k_{nl})$. However, the second-order scheme requires a stronger smoothing with $k_{gs}/k_{nl} = 1.2$ than the first-order scheme, which requires $k_{gs}/k_{nl} = 1.45$. But the second-order scheme has much less scatter in the optimum smoothing length for the different models than the first-order scheme.

The cross-correlation function $S(R_g)$ for different smoothing scales $R_g$ applied to the final density fields in real space is shown in Fig. 3. Here we measure the smoothing scale $R_g$ in units of $1/128^{\text{th}}$ of the simulations size, i.e. in grid units of the resolution of the initial data.

For the stage $z = 0$ according to COBE normalization of BSI initial data the cross-correlation coefficient indicates an excellent performance already on scales of the resolution limit ($R_g = 1$). This remains true for all three sizes of the simulations we have investigated, although for the smallest box ($75h^{-1}$Mpc) the performance of the analytical schemes is worse than for the two larger boxes. At earlier times ($z = 1$ and $z = 2$) the performance of the analytical schemes for the two smaller boxes ($75h^{-1}$Mpc and $200h^{-1}$Mpc) is better than for $z = 0$.

For COBE-normalized standard CDM the performance of the analytical schemes at time $z = 0$ is not as good as for BSI, for the $75h^{-1}$Mpc box we get an acceptable performance (comparable to that for BSI at $R_g = 1$) only on scales of $R_g \geq 2.5\ldots 3$. The performance on the larger boxes is comparable to that on the next smaller ones for BSI. At the earlier times ($z = 1$ and $z = 2$) the performance of the optimized perturbation schemes with standard CDM is not as good as for BSI initial data, in a way which is similar to the results for $z = 0$.

Due to the Gaussian smoothing applied to the initial data in Fourier space, the power-spectrum of the final density distribution for the optimized Lagrangian schemes (Figs. 4 and 5) is not even as high as that of the linearly evolved power-spectra, both for SCDM and BSI initial density distributions, though for the latter — due to its higher $k_{nl}$ — the Lagrangian approximations differ much less from the linearly evolved power spectrum, even at the late stage $z = 0$. Looking at the convolution of the linearly evolved power spectrum with the square of the window function (9) shows that the optimized Lagrangian schemes do not only produce more small-scale (large $k$) power than those without optimization, but that their power-spectra are at the small-scale end several orders of magnitude higher than expected for a linear evolution of the smoothed initial density distributions. This feature is especially prominent for the SCDM initial data, whose nonlinear evolution at each stage reaches larger length scales than that of the BSI initial data. In view of the fact that the convolved linear spectra nearly lie above each other one can also see that our procedure consistently yields comparable smoothing lengths in physical units, almost independent of the size of the simulation box.

The density distribution functions (Fig. 6) show an over-representation of lower density peaks and an under-representation of the highest density peaks for the optimized Lagrangian approximation schemes in comparison with the numerical simulations, consistently for both kinds of initial data and for each step in the temporal evolution. The density values where the density distribution functions of the Lagrangian schemes intersect with the density distribution functions of the numerical simulations decrease with earlier times. Consistently throughout all of our simulations, the density distribution functions of the second-order optimized Lagrangian perturbation theory lie closer to the numerical ones than those of the first-order scheme.

Finally, a look at the two-point correlation function at $z = 0$ (Fig. 7) confirms the high quality of the optimized Lagrangian perturbation theory. While the correlation function of the optimized schemes on scales below the correlation length $r_0$ ($\xi(r_0) = 1$) lies well below that of the numerical simulations, the correlation length itself is underestimated only slightly (by less than $1h^{-1}\text{Mpc}$) by the optimized Lagrangian schemes. For the largest boxes ($500h^{-1}\text{Mpc}$), the somewhat larger error in $r_0$ is mainly determined by the resolution limit of the initial data. Above the correlation length the correlation functions of optimized Lagrangian perturbation theory and numerical simulations come into good agreement. In contrast to this, the Lagrangian perturbation theory without a smoothing of the initial conditions yields a much too low estimate of the correlation length, especially for standard CDM initial data (having a higher $\xi(r)$ at small scales than the BSI initial data).

## 5 Conclusions

Overcoming the problem of runaway evolution of small-scale modes in Lagrangian perturbation models by filtering out these modes in the initial data seems, at first glance, counterproductive. However, earlier work as well as the results of this work show that a genuine *gain in performance on small scales* can be obtained by such an optimizing approach. These and the present investigations clearly show the following:

1. The smoothing scale in the initial data for a set of (in their amount of small-scale power) quite different models can be consistently connected to the scale of nonlinearity, eq. (10). Over a broad range of effective power indices $n(k_{nl})$, corresponding to different initial power spectra, different times in the gravitational evolution and different sizes of the simulations, the width of a Gaussian window (9) given in terms of the scale of nonlinearity $k_{nl}$, can be fitted to be nearly constant for varying $n(k_{nl})$ for both first- and second-order Lagrangian schemes. Here the second-order scheme shows much less scatter in $k_{gs}/k_{nl}(n)$ and requires somewhat larger smoothing lengths in real space than the first-order scheme. For the latter a Gaussian smoothing window of width $k_{gs}/k_{nl} = 1.45$ seems to lie close to the optimum for most of the models considered here, while for second-order $k_{gs}/k_{nl} = 1.2$ is close to the optimum for virtually all models and times considered here. This is consistent with the results of Melott *et al.* (1994,1995), obtained for an even broader range of $n(k_{nl})$. Thus, for a given initial power spectrum and time evolution, the optimal smoothing can be given by an absolute physical length scale and is independent of the scale of the simulations.

2. The performance of the optimized Lagrangian approximation schemes with respect to the *evolution of the density field* is excellent for larger boxes down to their resolution limit (see Fig. 8 for slices through our largest SCDM- and BSI-models). For smaller boxes, which resolve smaller physical scales, the approximation of the density field is satisfactorily modeled only on scales above the resolution limit: For SCDM initial data, the density cross-correlation coefficient at $z = 0$ shows that the Lagrangian schemes differ from the numerical simulations by about 2% on scales of $\sim 1.5\ldots 2.0 h^{-1}$Mpc, for BSI initial data this difference occurs on scales in the range $\sim 1.0\ldots 1.5 h^{-1}$Mpc.

3. The performance of the optimized Lagrangian approximation schemes with respect to the *clustering point process* is very good down to scales close to the correlation length $r_0$ of the two-point correlation function. The correlation length $r_0$ itself is underestimated at the time $z = 0$ by about 10-20 %, for BSI and SCDM. Especially for models with a high amount of small-scale power in the initial data, like SCDM, the Lagrangian schemes display a considerable gain in performance by introducing an optimized smoothing of the initial data.

4. The removal of small-scale power in the initial density fluctuation field actually causes the Lagrangian perturbation schemes to *produce more power on small scales*. Although optimized Lagrangian perturbation schemes still produce less power on small scales than predicted by Eulerian linear theory, they not only produce more power than Lagrangian perturbation theory without an optimized smoothing of the initial data (which is confirmed in an analytical calculation in Schneider & Bartelmann 1995), but also introduce large amounts of small-scale power compared to the linear evolution of their smoothed initial data.

These results clearly indicate, that at least in cases where one is interested in the formation of large-scale structure on scales above the correlation length, the optimized approximations considered here are a viable way to replace numerical simulations. This is especially true in cases when one is interested in the simulation of the large-scale distribution of matter as in the analysis of the cluster distribution, see e.g. Borgani *et al.* 1995, these authors have chosen for the first-order scheme a smoothing which uses a somewhat smaller physical smoothing length-scale, which is still near to the optimal one determined by this work. In this context we wish to emphasize that an increase of the smoothing length-scale, i.e. a decrease of $k_{gs}$, beyond the optimal one soon significantly decreases the performance of the Lagrangian perturbation schemes, while at least for negative $n(k_{nl})$ a decrease of the smoothing length-scale shows a much smaller effect. Numerical simulations, however, still remain the optimal tool in the range $1h^{-1}\text{Mpc}\ldots r_0$, both precision-wise and cost-wise. Only below this range hydrodynamic effects contribute significantly to the evolution of structures (see e.g. Frenk *et al.* 1995). An other interesting application of the Lagrangian perturbation schemes lies in the simulation of sparse surveys like, e.g., pencil-beams. Here the analyticity of the solution (1) allows running higher absolute particle densities by interpolation of the mapping (1). This higher particle density in comparison with numerical simulations of the same resolution is especially beneficent when introducing a high selection effect in the post-processing of the particle distribution, as is necessary for the simulation of galaxy surveys. For an example of the possibilities offered by such a procedure see Weiß & Buchert (1993). In addition to this it is notable that the time needed to realize the mapping (1) corresponds to roughly one time-step of a numerical particle-mesh simulation; the gain in execution time of the simulations is considerable. Thus, our analytical schemes provide a usefull tool for the simulation of large-scale and very large-scale observations as well as for a quick checking of, e.g., the large-scale behavior of newly constructed initial data.

## Acknowledgements

We wish to thank Adrian L. Melott (Univ. of Kansas) for valuable remarks as well as for the permission to use his programs for the crosscorrelation statistics, which were also used in Buchert *et al.* (1994) and Melott *et al.* (1995). AGW wishes to thank the AIP in Potsdam for the opportunity to work on this project during a stay at the AIP. TB acknowledges support of the Sonderforschungsbereich 375–95 für Astro-Teilchenphysik der Deutschen Forschungsgemeinschaft. SG wishes to thank the MPA in Garching for its hospitality.

# Figure captions

Figure 1: The linear power spectrum of the SCDM model (dashed line) and the BSI model (solid line).

Figure 2: Optimal smoothing length in terms of $k_{\rm nl}$ for SCDM and BSI spectra.
Top: first-order optimized Lagrangian perturbation theory (OLPT1); bottom: second-order scheme (OLPT2).
SCDM: triangles $\div$ $z=0$; asterisks $\div$ $z=1$; crosses $\div$ $z=2$.
BSI: diamonds $\div$ $z=0$; +-sign $\div$ $z=1$; for $z=2$ $k_{nl}$ lies outside the box.

Figure 3: Density cross-correlations for SCDM and BSI.
dash dotted lines: OLPT1; dashed lines: OLPT2.
Thick lines: $500h^{-1}$Mpc; medium lines: $200h^{-1}$Mpc; thin lines: $75h^{-1}$Mpc boxes.

Figure 4: Power spectra for SCDM and BSI at $z=0$.
Thin solid lines $\div$ linear spectrum; lower one smoothed with optimum $2^{nd}$ order window.
Thick solid lines $\div$ numerical simulation.
Dash-dotted lines $\div$ first order; dashed lines second order.
Thick lines $\div$ OLPT; thin lines $\div$ Lagrangian perturbation theory without optimization (LPT).
Vertical bars denote the Nyquist frequency of the initial data.
The dotted vertical line shows the nonlinearity scale $k_{nl}$.

Figure 5: The same as in Figure 4, for times $z=1$ and $z=2$.

Figure 6: Density distribution functions for SCDM and BSI.
Solid lines $\div$ numerical simulations; dash-dotted lines $\div$ OLPT1; dashed lines $\div$ OLPT2.
Thick lines $\div$ $500h^{-1}$Mpc; medium lines $\div$ $200h^{-1}$Mpc; thin lines $\div$ $75h^{-1}$Mpc boxes.

Figure 7: Two-point correlation function for $z=0$ SCDM and BSI.
Solid lines $\div$ numerical simulations; dash-dotted lines $\div$ first order; dashed lines $\div$ second order.
Thin lines: LPT; thick lines: OLPT.

Figure 8: A slice through our $500h^{-1}$Mpc SCDM (left) and BSI (right) simulations.
Top: numerical simulation; bottom: OLPT2.

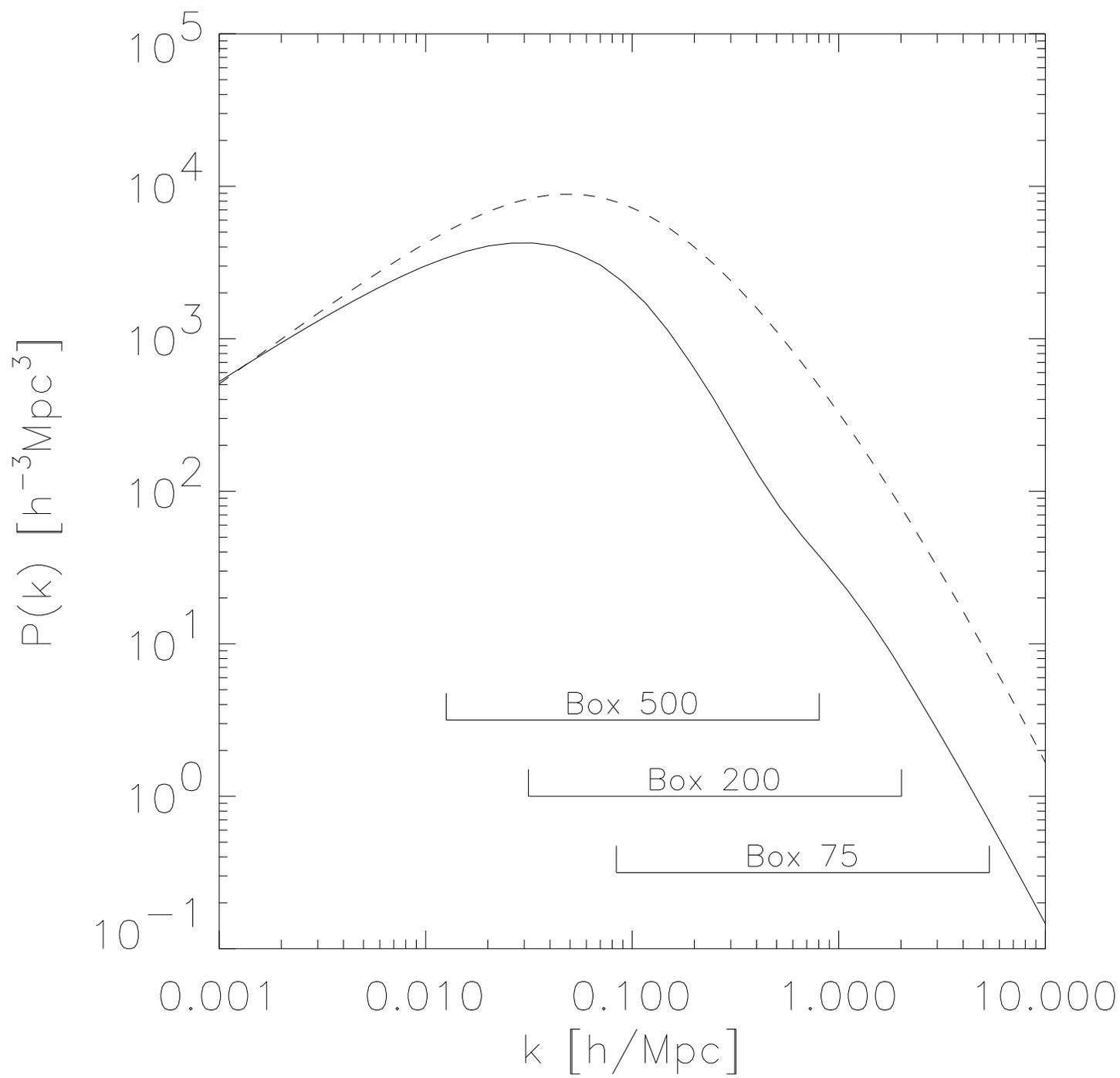

Figure 1: Initial power spectra linearly transported to $z = 0$

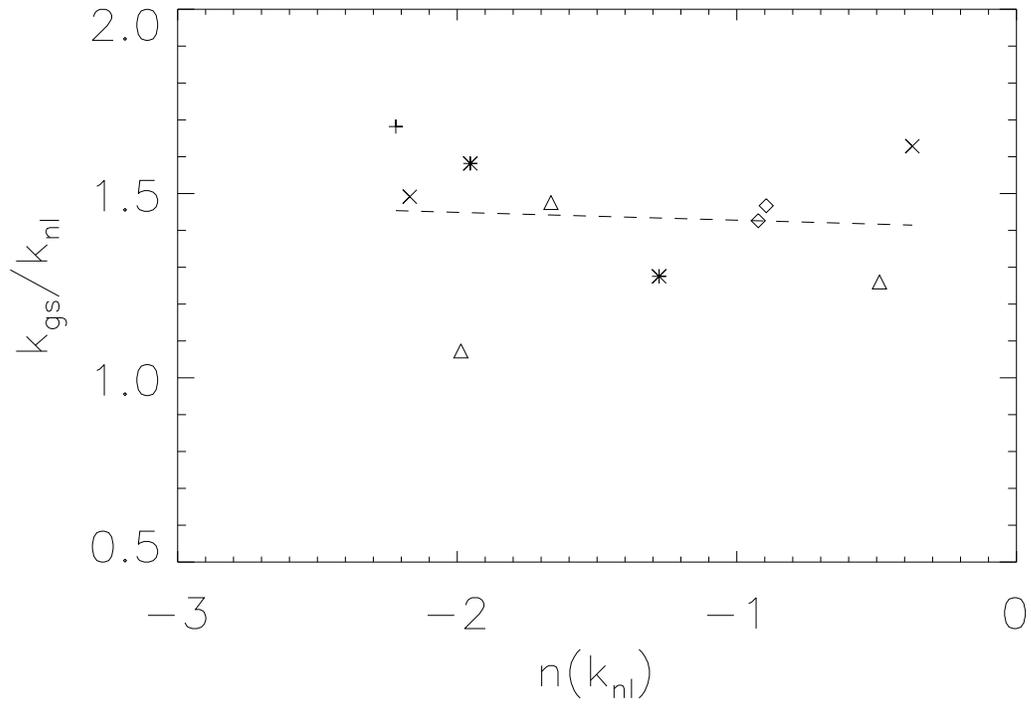

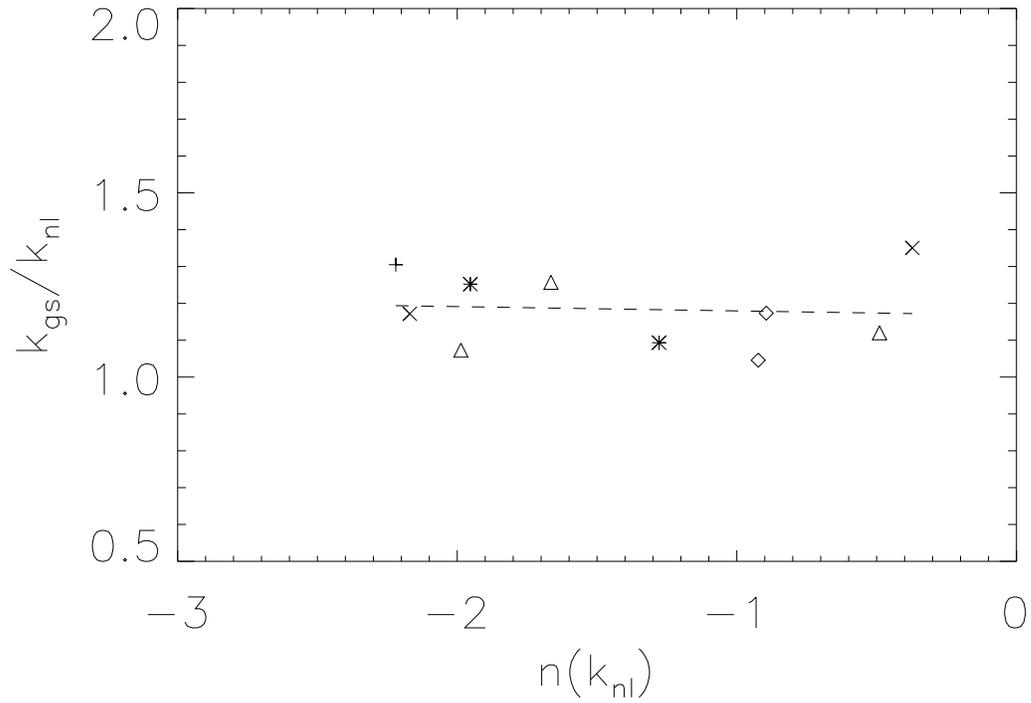

Figure 2: Smoothing scales

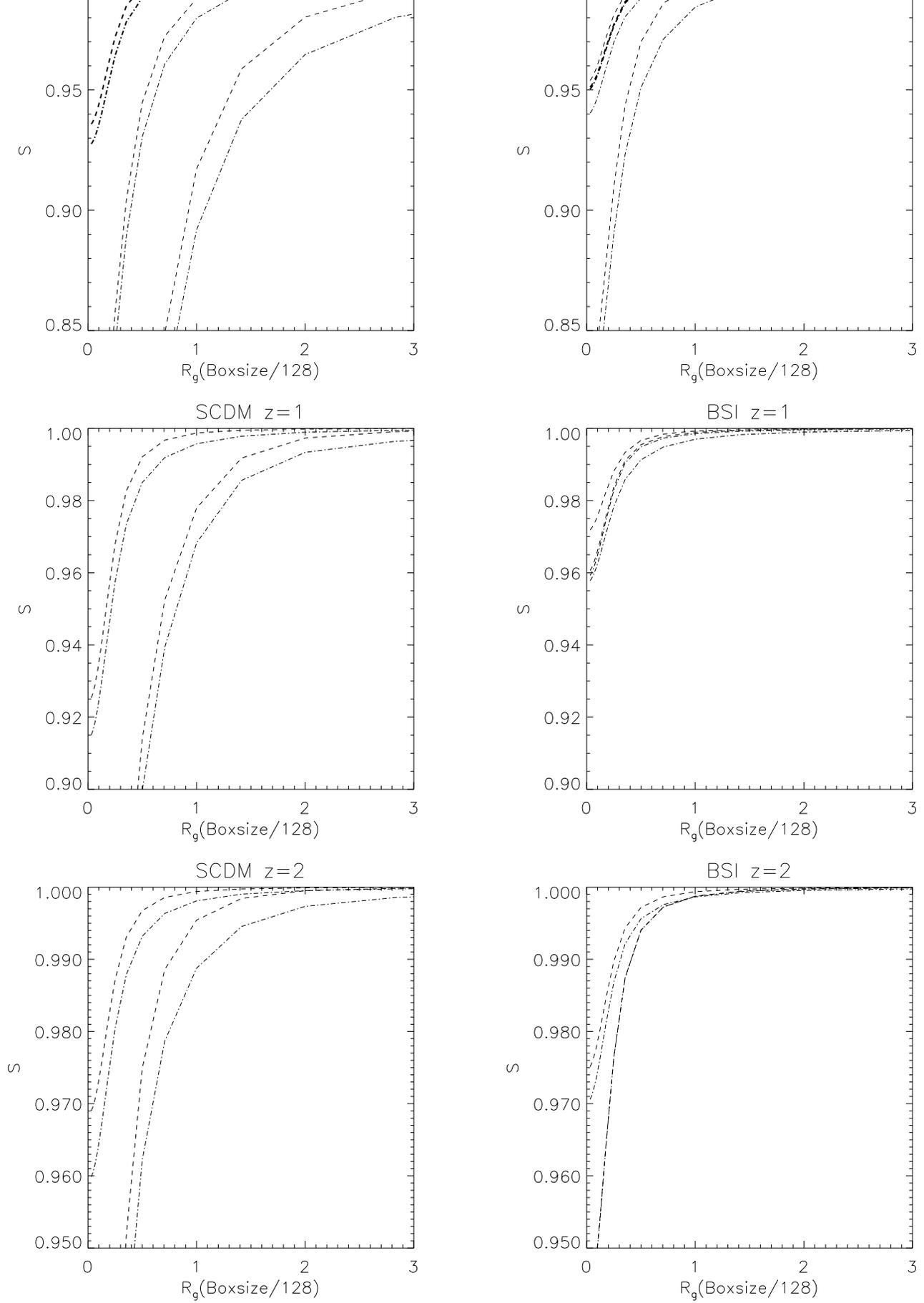

Figure 3: Cross correlations

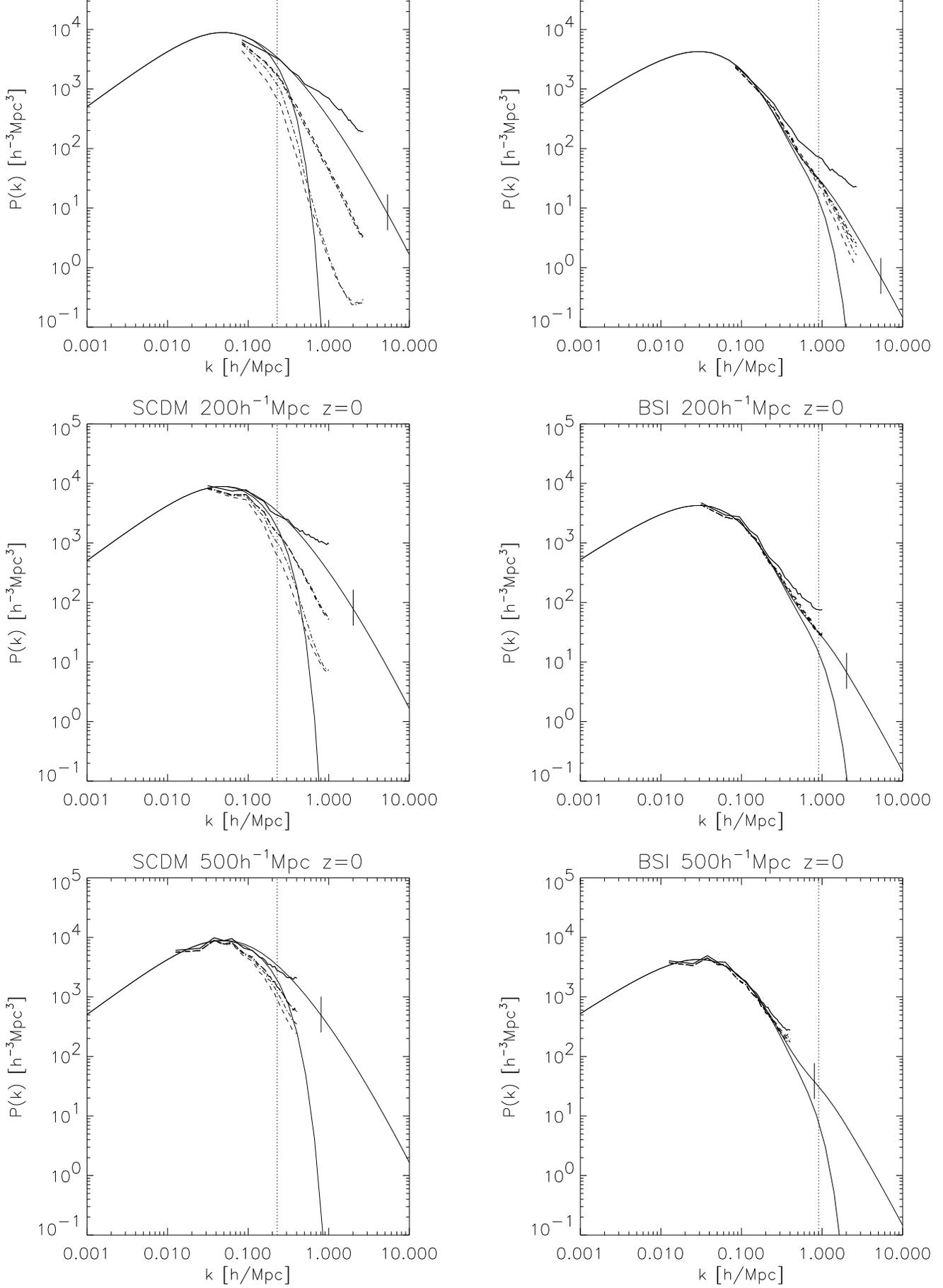

Figure 4: Power spectra at $z = 0$

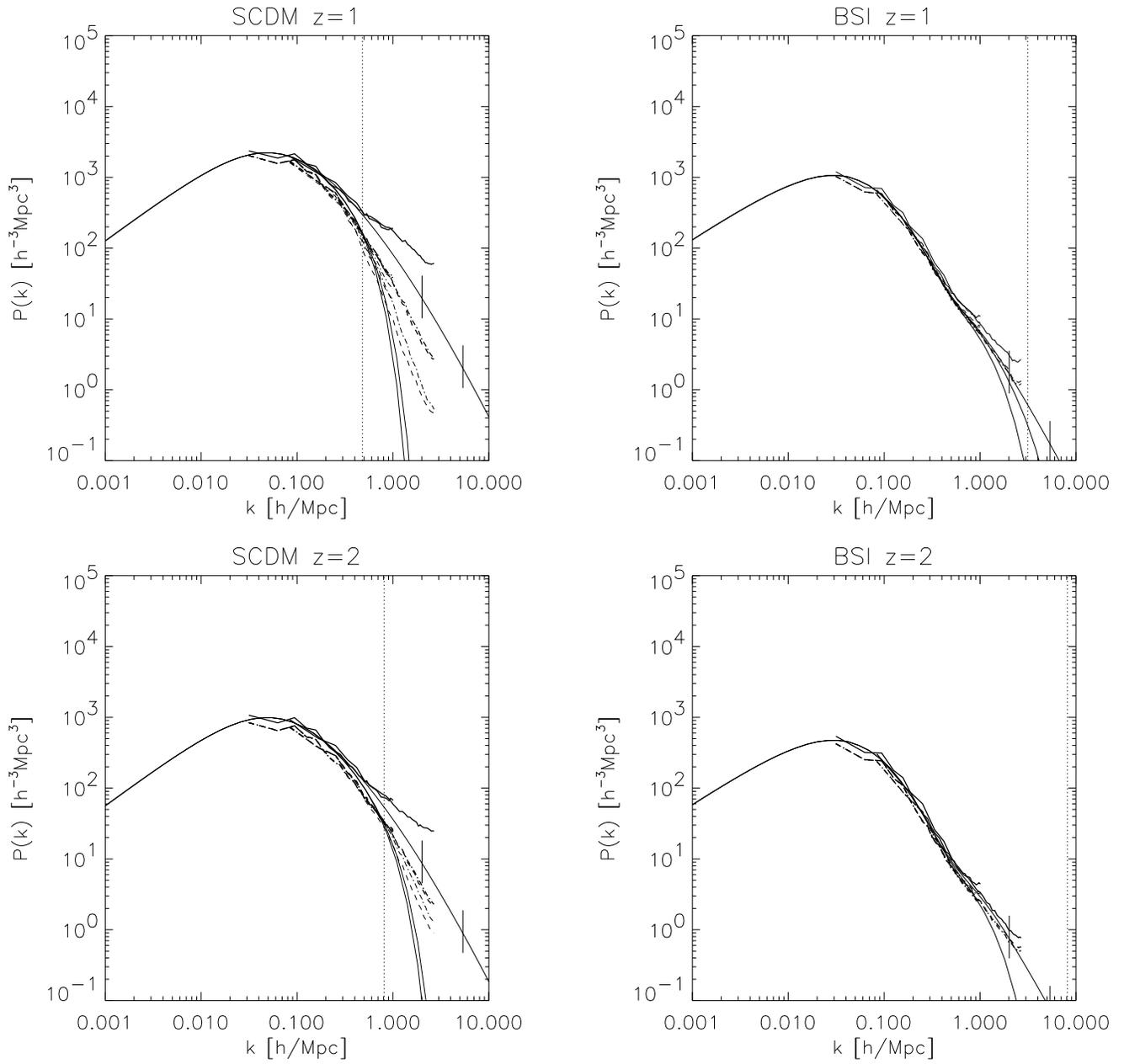

Figure 5: Power spectra at $z = 1$ and $z = 2$

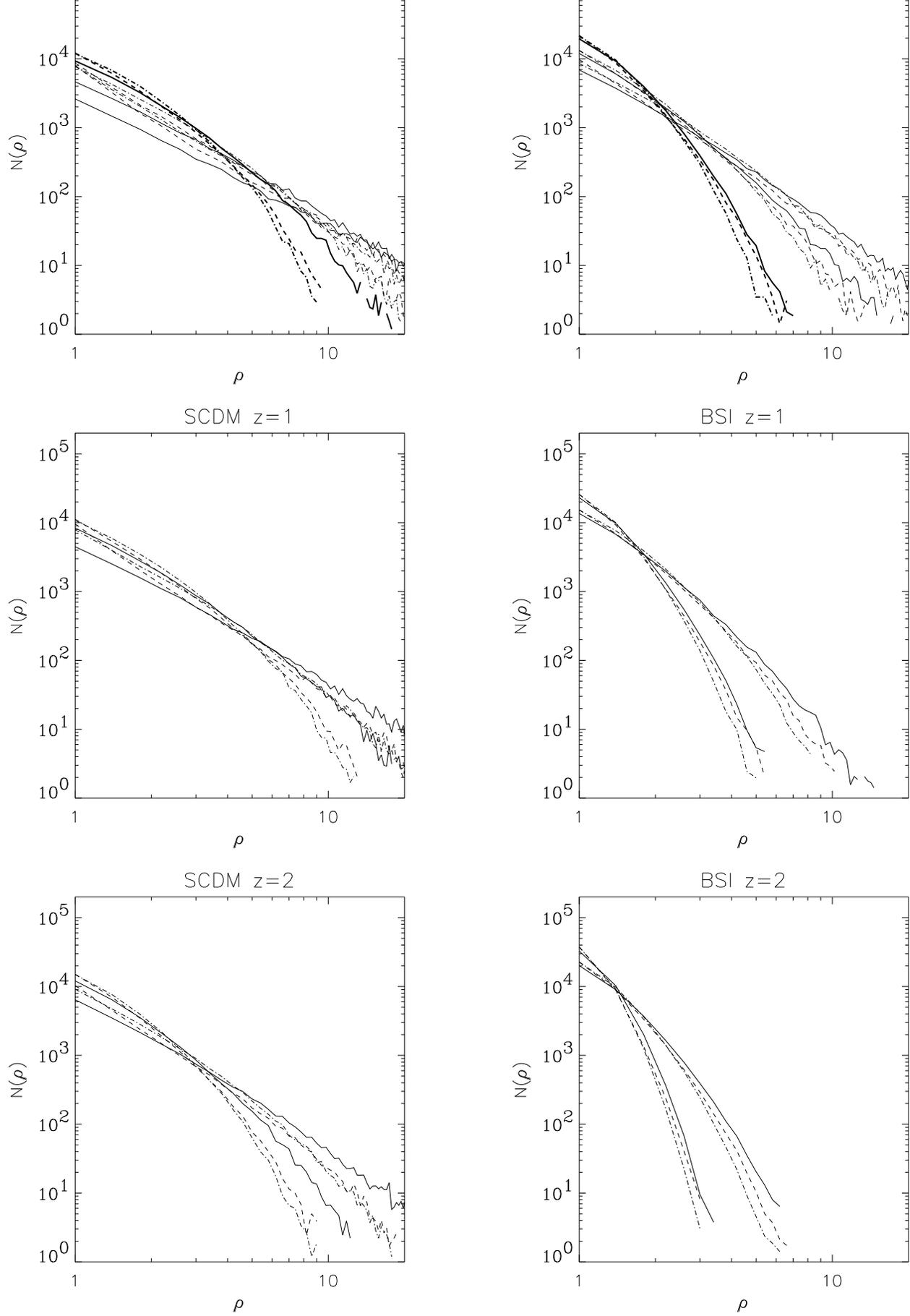

Figure 6: Density distribution functions

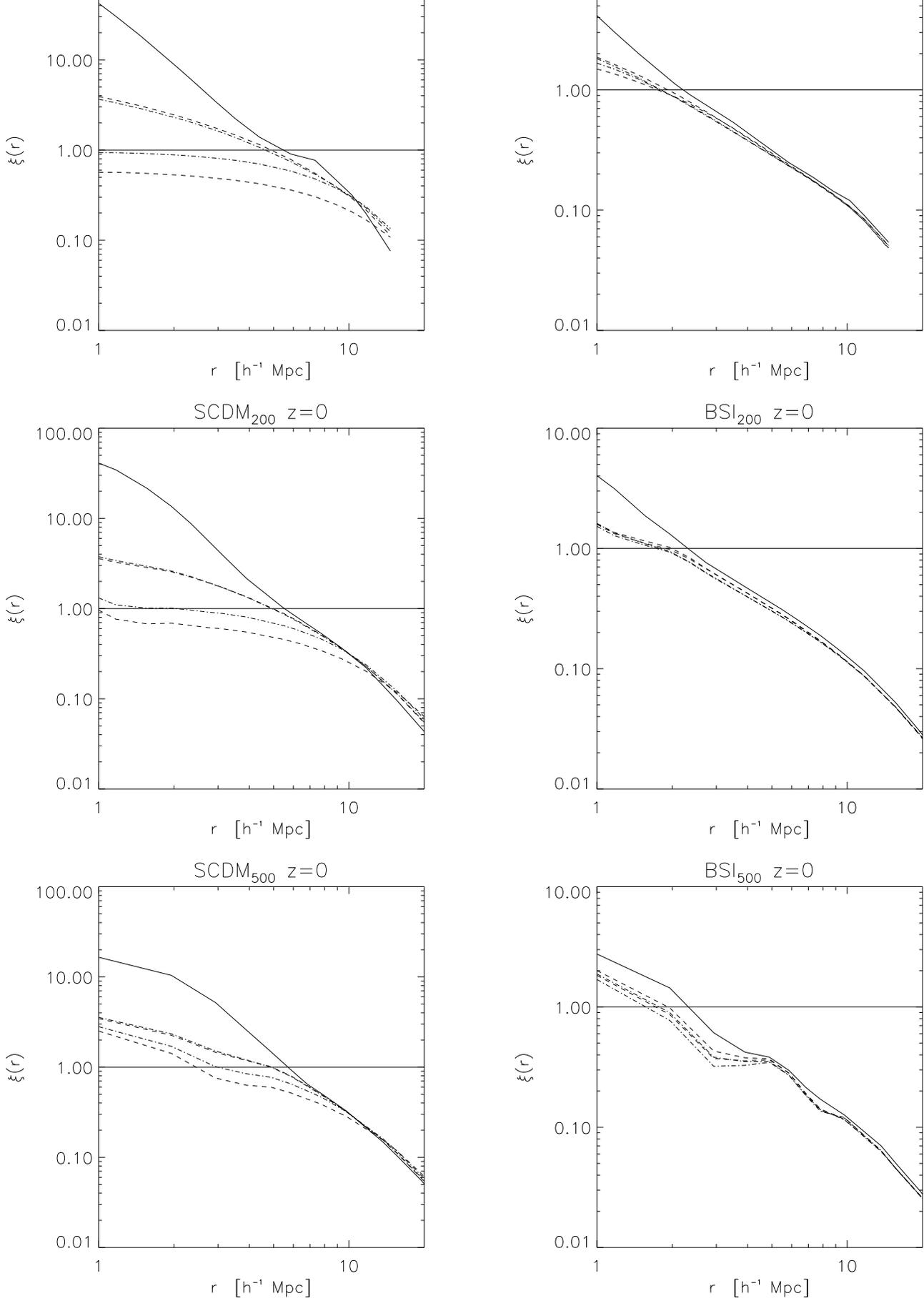

Figure 7: Two-point correlation functions